\documentclass[fleqn,twoside,fps]{article}
\usepackage{espcrc2,graphicx,latexsym,amssymb,amsmath}
\usepackage[figuresright]{rotating}

\title{Comparing iterative methods for overlap and twisted mass
  fermions\thanks{Poster presented by K.J.}
}

\author{
T.~Chiarappa\address[NICDESY]{ 
    NIC/DESY Zeuthen, Platanenallee 6,
    D--15738 Zeuthen, Germany}, 
K.~Jansen\addressmark[NICDESY],
K.-I.~Nagai\addressmark[NICDESY], 
M.~Papinutto\addressmark[NICDESY],
L.~Scorzato\address[HUB]{
    Humboldt Universit\"at zu Berlin, Institut f\"ur
    Physik, Newtonstrasse 15, D--12489 Berlin, Germany},
A.~Shindler\addressmark[NICDESY],
C.~Urbach\addressmark[NICDESY]\address[FUB]{
    Freie Universit\"at Berlin, Institut f\"ur Theoretische Physik, 
    Arnimallee 14, D--14196 Berlin, Germany}, 
U.~Wenger\addressmark[NICDESY], 
I.~Wetzorke\addressmark[NICDESY]
}

\begin{document}

\begin{abstract}
  We present a systematic comparison of various iterative methods to obtain
  the fermion propagator with both overlap and twisted mass fermions at fixed
  pion mass in the quenched approximation. Taking the best available algorithm
  in each case we find that calculations with the overlap operator are by a
  factor of 20-40 more expensive than with the twisted mass operator at the
  parameter values considered here.  For the overlap operator we also compare
  the efficiency of various methods for calculating the topological index.
\end{abstract}

\maketitle

\section{INTRODUCTION}

The Wilson twisted mass (Wtm) lattice fermion action of an SU$_f(2)$ flavour
doublet of mass degenerate quarks with maximal twist angle has the form
\begin{equation}
  D_{\text{tm}}(\mu) = D_\text{w}(m_\text{tm}) + 
  i \gamma_5  \tau_3 \cdot \mu \nonumber
\end{equation}
where $D_\text{w}$ is the Wilson Dirac operator, $m_\text{tm}$ the bare quark
mass tuned to its critical value and $\mu$ the twisted quark mass.  For our
purpose it is sufficient to consider only one of the two flavours.

The massive overlap operator is defined as
\begin{equation}
  D(m_\text{ov}) = 
  \Big(1 - \frac{m_\text{ov}}{2 \rho}\Big) D + m_\text{ov} \nonumber
\end{equation}
where 
\begin{equation}
  D = \rho \Big(1 + 
  \gamma_5 \textrm{sign}\big(Q(\rho)\big)\Big)  \nonumber
\end{equation}
is the massless overlap operator with $Q(\rho) = \gamma_5 D_\text{W}(-\rho)$,
$\rho=1.6$ and $m_\text{ov}$ the bare quark mass.

For both Wtm and the overlap operator we present results using the inversion
algorithms GMRES(m), CG(NE), CGS and BiCGstab \cite{Saad} and additionally MR
\cite{Saad} and SUMR \cite{Jagels,Arnold:2003sx} for the overlap operator.

\section{SETUP}
Our set-up consists of two quenched ensembles of 20 lattices with $V=12^4$ and
$16^4$ each generated with the Wilson gauge action at $\beta=5.85$
corresponding to a lattice spacing of $a \sim 0.12$ fm.

We invert both the twisted mass and the overlap operator on two point-like
sources $\eta$ with two different bare quark masses and require a stopping
criterion $|Ax-\eta|^2 < 10^{-14}$.

The quark masses $m_\text{ov}=0.10$ and $m_\text{ov}=0.03$ for the overlap and
$\mu=0.042$ and $\mu=0.0125$ for the twisted mass are chosen such that the
corresponding pion mass for the twisted mass and the overlap operator are
matched:
\begin{equation}
  \begin{array}{lcl}
    m_\pi = 720 \text{MeV}  &\Rightarrow&  \left\{ \begin{array}{rcl}
      m_\text{ov}&=&0.10   \\
      \mu&=&0.042  \nonumber
    \end{array} \right.\\
    && \\
    m_\pi = 390 \text{MeV} &\Rightarrow&  \left\{ \begin{array}{rcl}
      m_\text{ov}&=&0.03    \\
      \mu&=&0.0125 \nonumber
    \end{array} \right.
  \end{array}
\end{equation}

We are working in a chiral basis and the two sources are chosen so that they
correspond to sources in the two different chiral sectors.  For the CG(NE)
algorithm using the overlap operator we can then use the relation $P_\pm
D(m_\text{ov}) D(m_\text{ov})^\dagger P_\pm = 2 \rho P_\pm
D(m_\text{ov}^2/(2\rho)) P_\pm$ where $P_\pm$ are the chiral projectors, since
the inversions take place in a given chiral sector. This saves a factor of two
with respect to the general CG(NE) case and in the following we denote this
algorithm by CG$_\chi$.

The computational bottleneck for the inversions of the overlap operator is the
computation of the approximation of the sign-function $\text{sign}(Q)$. Our
approximations use Chebyshev polynomials of the order $O(200-300)$. In order
to achieve this we project out the lowest 20 and 40 eigenvectors of the
hermitian Wilson Dirac operator $Q(\rho)$ on the $12^4$ and $16^4$ lattice,
respectively.

It is well known that by adapting the accuracy of the approximation during the
course of the iteration one can speed up the inversions by large factors since
a reduction in the order of the polynomial enters multiplicatively in the
total cost of the inversion. In the following we denote these algorithms by
the subscript $_\text{ap}$ for adaptive precision.  The precision is adapted
so as to ensure that no contributions to the sign-function approximation are
calculated which are not needed at the present stage of the algorithm.  In the
case of the CG$_\text{ap}$ we simply calculate contributions up to the point
where they are smaller by a factor $O(10^{-2})$ than the desired residuum.
This requires the full polynomial only at the beginning of the CG-search while
towards the end of the search we use polynomials of the order $O(10)$.  In the
case of MR$_\text{ap}$, we start with a low order approximation of $O(10)$
right from the beginning. Subsequently the introduced error is corrected from
time to time by calculating the true residuum to full precision.

\begin{table}[thb]
  \begin{tabular}{|l||c|c|c|}
    \hline
 $V,m_\pi$   &  Overlap  & Wtm & rel. factor \\

    \hline
    $12^4, 720$Mev & 48.8(6)   & 2.6(1) & 18.8 \\
    $12^4, 390$Mev & 142(2)    & 4.0(1) & 35.4 \\
    \hline
    $16^4, 720$Mev & 225(2)    & 9.0(2) & 25.0 \\
    $16^4, 390$Mev & 653(6)    &17.5(6) & 37.3 \\
    \hline
  \end{tabular}\\[2pt]
\caption{Best absolute timings in seconds.}
\label{tab:absolute timings}
\end{table} 

\section{RESULTS}
\begin{figure*}[thb]
\begin{center}
\vspace{-1cm}
\begin{eqnarray*}
\includegraphics[height=5.5cm]{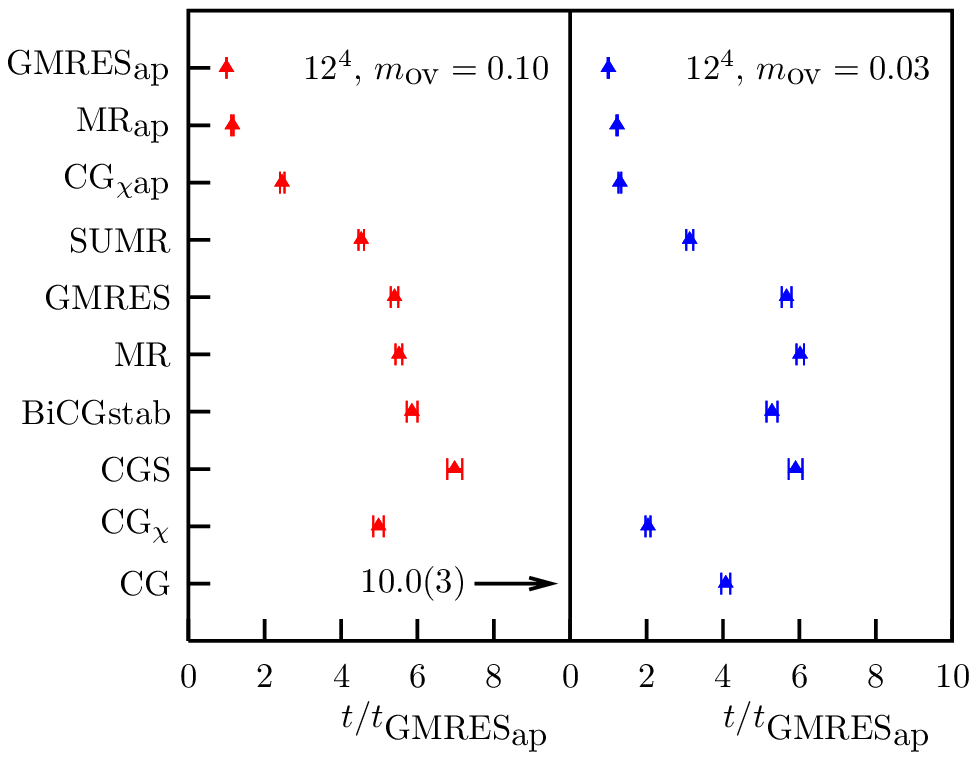} &
\includegraphics[height=5.5cm]{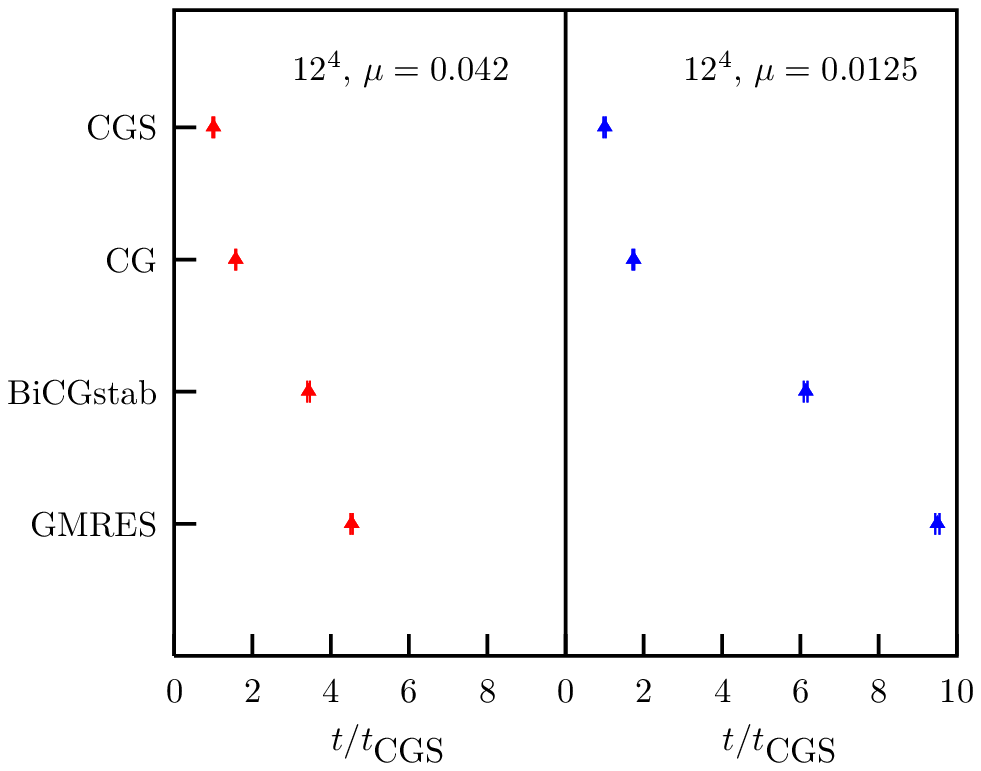} \\
\includegraphics[height=5.5cm]{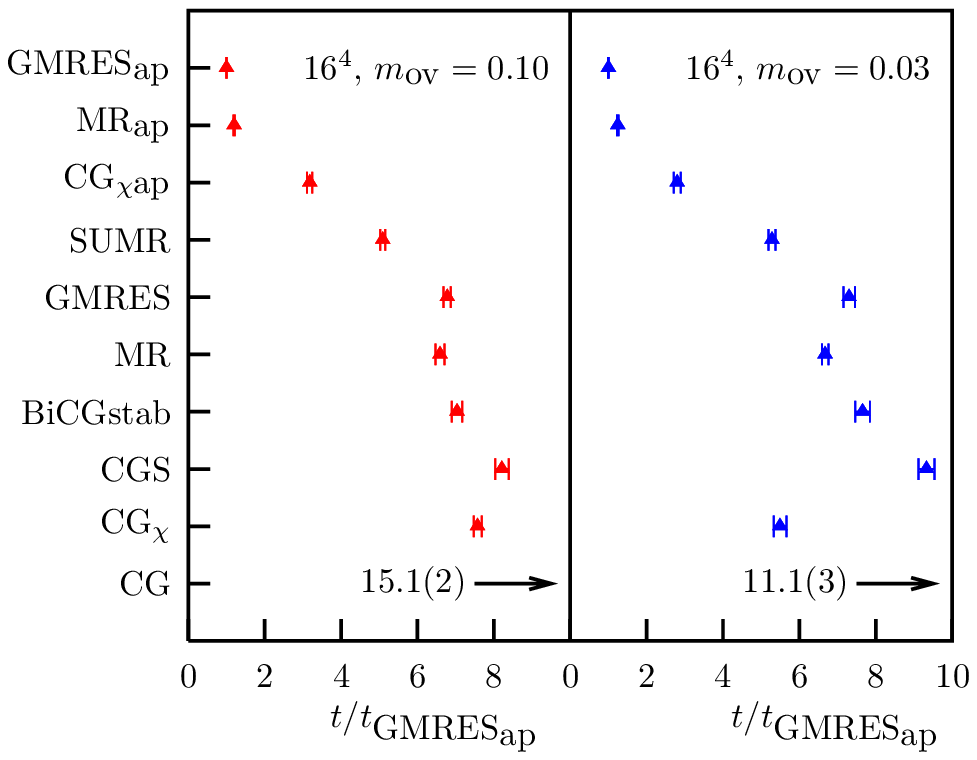} &
\includegraphics[height=5.5cm]{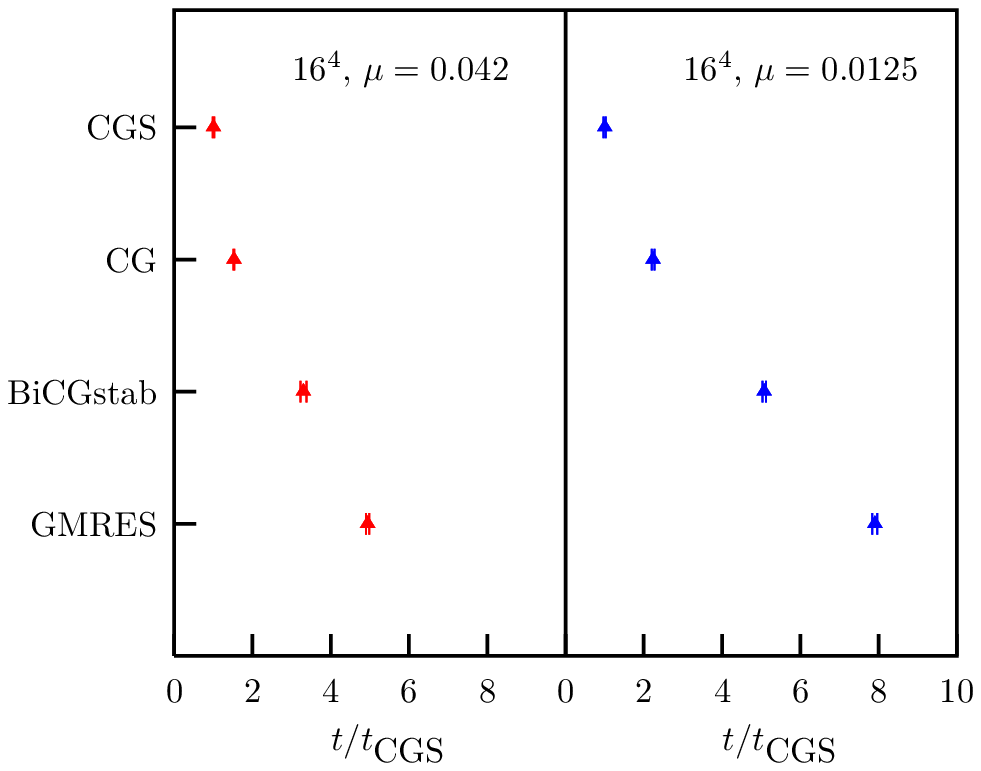} 
\end{eqnarray*}
\end{center}
\vspace{-1.5cm}
\caption{{}Relative timings of the various algorithms for the overlap
(left column) and the twisted mass operator (right column).}
\vspace{-0.2cm}
\label{fig:relative timings}
\end{figure*}
We first present the relative timings of the algorithms for both the overlap
and twisted mass operators in Figure \ref{fig:relative timings} where in each
case the timings are relative to the fastest algorithm. The timings were all
obtained on one node of the Juelich Multiprocessor (JUMP) IBM p690 Regatta
using 32 processors.

Next we compare directly the absolute and relative cost for the overlap and
twisted mass operator in Table \ref{tab:absolute timings} where we pick in
each case the best available algorithm, GMRES$_\text{ap}$ for the overlap and
CGS for twisted mass.

\section{TOPOLOGICAL CHARGE COMPUTATION}
For the computation of the topological index it is important to note that the
determination of the chiral sector which contains the zero-modes comes for
free when one uses the CG-algorithm for the inversion.  From the
CG-coefficients which are obtained during the iteration one can build up a
tridiagonal matrix related to the underlying Lanczos procedure. The
eigenvalues of this matrix approximate the extremal eigenvalues of the
operator and it turns out that the lowest 5-10 eigenvalues are approximated
rather accurately. By estimating the eigenvalues once in each chiral sector
and by pairing them accordingly it is possible to identify the chiral sector
which contains zero modes.
               
In order to determine the topological charge itself one has to compute the
lowest eigenvalues of the overlap operator as well as their degeneracies.  We
have implemented two different algorithms, one based on the Ritz-Jacobi (RJ)
method \cite{Kalkreuter:1995mm,Giusti:2002sm} and the other based on the
Jacobi-Davidson (JD) method \cite{Sleijpen}.  Both of them are improved by
looking separately in the two chiral sectors using adaptive precision.

We compare the two algorithms on a $12^3\times24$ and a $16^4$ lattice at
$\beta=5.85$ and $\rho=1.6$ with five configurations each. Both methods first
determine the chiral sector containing zero modes and subsequently all the
zero modes are calculated in this sector. With the JD method additionally two
non zero modes have been computed. The timings relative to RJ are 0.73(9) on
the $12^3\times24$ lattice and $0.93(7)$ on the $16^4$ lattice.

The performance of both methods is comparable with slight advantages for the
JD method.  The speedup for the two methods in a MPI-parallel program is
equal.

\section{CONCLUSIONS AND OUTLOOK} 
We find that for Wtm fermions CGS appears to be the fastest inversion
algorithm while for overlap fermions it is GMRES$_\text{ap}$ for the
parameters investigated here. In a direct competition between twisted mass and
overlap fermions the latter are by a factor of 20-40 more expensive if one
compares the best available algorithms in each case.

For the index computation a clever combination of the methods described above
looks most promising.  Finally we note that one can apply various kinds of
preconditioning to all the algorithms investigated here. For the twisted mass
operator we expect even/odd or SSOR preconditioning to be efficient, while for
the overlap operator low-mode preconditioning should be very effective for low
quark masses.



\begin{thebibliography}{9}

\bibitem{Saad} 
  Y.~Saad, {\it Iterative Methods for sparse linear
  systems}, 2nd ed., SIAM, 2003.

\bibitem{Jagels} 
  C.F.~Jagels and L.~Reichel, 
  Numer.~Linear
  Algebra Appl. 1(6), 555-570 (1994).
                    
\bibitem{Arnold:2003sx}
G.~Arnold, N.~Cundy, J.~van den Eshof, A.~Frommer, S.~Krieg, T.~Lippert and K.~Sch\"afer,
arXiv:hep-lat/0311025.

\bibitem{Kalkreuter:1995mm}
T.~Kalkreuter and H.~Simma,
Comput.\ Phys.\ Commun.\ {\bf 93}, 33 (1996) [arXiv:hep-lat/9507023].

\bibitem{Giusti:2002sm}
L.~Giusti, C.~Hoelbling, M.~L\"uscher and H.~Wittig,
Comput.\ Phys.\ Commun.\  {\bf 153} (2003) 31
[arXiv:hep-lat/0212012].

\bibitem{Sleijpen} 
  G.L.G.~Sleijpen and H.A.~Van der Vorst, 
  J.~Matrix Analysis Appl. 17, 401-425(1996).
     
\end{thebibliography}
\end{document}